\documentclass[oneside,onecolumn, aps,showpacs, showkeys]{revtex4}
\usepackage{graphicx}
\usepackage{amsmath, amsthm, amsfonts,amssymb}
\usepackage{graphicx}

\begin{document}
\title[]{Market Efficiency in Foreign Exchange Markets}

\author{Gabjin \surname{Oh}}
\email{gq478051@postech.ac.kr} \affiliation{Asia Pacific Center for Theoretical Physics \& NCSL, Department of Physics, Pohang University of Science
and Technology, Pohang, Gyeongbuk, 790-784, Korea}

\author{Seunghwan \surname{Kim}}
\email{swan@postech.ac.kr} \affiliation{Asia Pacific Center for Theoretical Physics \& NCSL, Department of Physics, Pohang University of Science
and Technology, Pohang, Gyeongbuk, 790-784, Korea}

\author{Cheoljun \surname{Eom}}
\email{shunter@pusan.ac.kr}\affiliation{Division of Business Administration, Pusan National University, Busan 609-735, Korea}

\received{13 11 2006}

\begin{abstract}

We investigate the relative market efficiency in financial market data, using 
the approximate entropy(ApEn) method for a quantification of randomness in time series. 
We used the global foreign exchange market indices for 17 countries during two periods from 1984 to 1998 and from 1999 to 2004 
in order to study the efficiency of various foreign exchange markets around the market crisis. 
We found that on average, the ApEn values for European and 
North American foreign exchange markets are larger than those for African and Asian ones except Japan. 
We also found that the ApEn for Asian markets increase significantly after the Asian currency crisis. 
Our results suggest that the markets with a larger liquidity such as European and North American foreign exchange markets 
have a higher market efficiency than those with a smaller liquidity such as the African and Asian ones except Japan.


\end{abstract}

\maketitle
\section{INTRODUCTION}

Recently, the complex features of financial time series have been studied using a variety of methods developed in econophysics [1-2]. 
The analysis of extensive financial data has empirically pointed to the breakdown of the efficient market hypothesis(EMH), in particular, the weak-form of EMH [4-6, 12-14]. 
For example, the distribution function of the returns of various financial time series 
is found to follow a universal power law distribution with varying exponents [4-6, 13]. 
The returns of financial time series without apparent long-term memory are found to possess the long-term memory in absolute value 
series, indicating a long-term memory in the volatility of financial time series [7,8,9,11,15]. 

In this paper, we use a method developed in statistical physics to test the market efficiency 
of the financial time series. The approximate entropy(ApEn) proposed by Pincus {\em et al.} can be used to quantify the randomness in the time series [16, 17]. 
The ApEn can not only quantify the randomness in financial time series 
with a relatively small number of data but also be used as a measure for the stability of time series \cite{Pincus2004}. Previously, the Hurst exponent 
was used to analyze various global financial time series, which suggested that the mature markets have features different from the emerging markets. 
Thus, the Hurst exponents for the mature markets exhibit a short-term memory, while those for the emerging markets exhibit a long-term memory [9, 12]. 
It was also shown that the liquidity and market capitalization may play an important role in understanding the market efficiency [10]. 
Using the ApEn, we study the market efficiency of the global foreign exchange markets. We use the daily foreign exchange rates for 17 countries from 1984 to 1998, 
and ones for 17 countries from 1999 to 2004 around the Asian currency crisis. 

We found that the ApEn values for European and North American foreign exchange markets are larger than those for  
African and Asian ones except Japan. 
We also found that the market efficiency of Asian foreign exchange markets measured by ApEn increases significantly after the Asian currency crisis. 

In Section \ref{sec:METHODOLOGY}, we describe the financial data used in this paper and introduce the ApEn method. In Section \ref{sec:RESULTS}, 
we apply the ApEn method to global foreign exchange rates and 
investigate the relative efficiency of the diverse foreign exchange markets. 
Finally, we end with a summary.

\section{DATA and METHOD}
\label{sec:METHODOLOGY}

\subsection{DATA}

We investigate the market efficiency of the financial time series for various foreign exchange markets. For this purpose, we use the return 
series of daily foreign exchange rates for 17 countries from 1984 to 1998 (Data A) and from 1999 to 2004 (Data B). 
The Data A and Data B are obtained before and after the Asian crisis, respectively. 
The data are grouped into European, North American, African, Asian and Pacific countries (from  http://www.federalreserve.gov/RELEASES/). 
The returns of the financial time series are calculated by a log-difference and properly normalized, respectively. 
The normalized return $R_{t}$ at a given time t is defined by 

\begin{equation}\label{e1}
{r_{t}}=   ln~ {P_{t}} - ln~ {P_{t-1}}, ~ \\
{R_{t}}= \frac {ln~ {P_{t}} - ln~ {P_{t-1}}}{\sigma(r_{t})}, \\
\end{equation}
where $P_{t}$ is the daily foreign exchange rate time series, $r_{t}$ the return time series after a log-difference, and 
$\sigma(r_{t})$ the standard deviation of the return.

\subsection{Approximate Entropy(ApEn)}

Pincus {\em et al.} proposed the ApEn to quantify the randomness inherent in 
time series data [16, 17]. Recently, Pincus and Kalman applied the ApEn method to a variety of financial time series 
in order to investigate various features of the market, in particular, the randomness [18]. The ApEn is defined as follows:

\begin{equation}\label{e2}
ApEn(m,r) =  \Phi^{m}(r) - \Phi^{m+1}(r),  \\
\end{equation}
where $m$ is the embedding dimension, $r$ the tolerance in similarity. The function $\Phi^{m}(r)$ is given by 

\begin{equation}\label{e3}
\Phi^{m} (r) = (N-m+1)^{-1} \displaystyle \sum_{i=1}^{N-m+1} \ln[C_{i}^{m}(r)], \\
\end{equation}

\begin{equation}\label{e4}
C_{i}^{m} (r) = \frac{\displaystyle B_{i}(r)}{(N-m+1)}, \\
\end{equation}

where $B_{i}(r)$ is the number of data pairs within a distance $r$, 

\begin{equation}\label{e5}
B_{i} \equiv d[x(i), x(j)] \leq r. 
\end{equation}
The distance $d[x(i),x(j)]$ between two vectors $x(i)$ and $x(j)$ in $R^{m}$ is defined by 

\begin{equation}\label{e6}
d[x(i),x(j)]= \underset{k=1,2,..,m}{max}(|u(i+k-1)-u(j+k-1)|),
\end{equation}
where $u(k)$ is a time series. 

The ApEn value compares the relative magnitude between repeated pattern occurrences for the embedding dimensions, $m$ and $m+1$. 
When the time series data have a high degree of randomness, 
the ApEn is large. On the other hand, ApEn is small for the time series with a low degree of randomness. 
Therefore, the ApEn can be used as a measure of the market efficiency. In this work, the ApEn is estimated with the embedding 
dimension, $m=2$, and the distance, $r=20\%$ of the standard deviation of the time series, similar to the preivous works[18].

\section{RESULTS}
\label{sec:RESULTS}

\begin{figure}[tb]
\includegraphics[height=12cm, width=15cm] {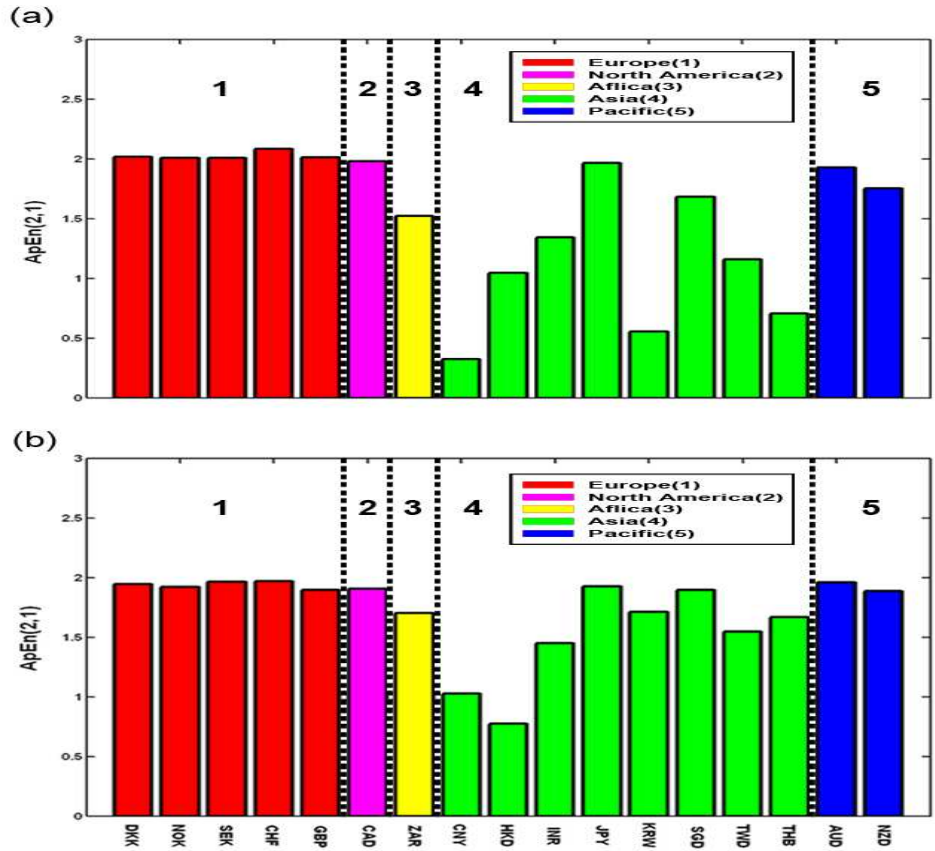}

\caption[0]{(a) ApEn for foreign exchange rates of 17 countries from 1984 to 1998 and (b) the ApEn for 
foreign exchange rates of 17 countries from 1999 to 2004. [1:Denmark (DKK/USD), 2:Norway (NOK/USD), 
3:Sweden (SEK/USD), 4:Switzerland (CHF/USD), 5:United Kingdom (GBP/USD), 6:Canada (CAD/USD), 7:South Africa (ZAR/USD), 8:China (CNY/USD), 9:HongKong (HKD/USD), 
10:India (INR/USD), 11:Japan (JPY/USD), 12:Korea (KRW/USD), 13:Singapore (SGD/USD), 14:Taiwan(TWD/USD), 15:Thailand (THB/USD), 16:Australia (AUD/USD), 17:New Zealand (NZD/USD)]. 
The red, pink, yellow, green and blue color bars correspond to European, North American, African, Asian and Pacific countries, respectively.}
\end{figure}

In this section, we investigate the relative market efficiency for various foreign exchange markets. We measure the randomness in financial 
time series using the approximate entropy (ApEn) method.
We analyze the ApEn values for the global foreign exchange rates in the Data A and Data B defined in Section II. 
Figure 1(a) shows the ApEn values for the foreign exchange rates of Data A before the Asian currency crisis. 
The red, pink, yellow, green, and blue colors denote the European, North American, African, Asian, and Pacific foreign exchange markets, respectively.
We found that the average ApEn for European foreign exchange rates is 2.0 and the ApEn for North American one is 1.98, which are larger 
than the ApEn values for Asian ones with 1.1 (except Japan), and African ones with 1.52. The ApEn for the Pacific foreign exchange rates is 1.84, 
which is intermediate between two groups. 
This is due to the liquidity or trading volumes 
in European and North American foreign exchange markets, which are much larger than those for other foreign exchange markets. 
The market with a larger liquidity such as European and North American foreign exchange markets 
shows a higher market efficiency than the market with a smaller liquidity such as Asian (except Japan) and African foreign exchange markets.
In order to estimate the change in market efficiency after the market crisis, we investigate the ApEn for the foreign exchange rates of Data B after the Asian currency crisis. 
Figure 1(b) shows the ApEn values for Data B. 
We found that the average ApEn values for European and North American foreign exchange markets 
do not change much from the case of Data A. However, the ApEn values for Asian increased sharply from 1.1 to 1.5 after the Asian currency crisis, 
indicating the improved market efficiency. 
Note that the ApEn of Pacific foreign exchange rates is 1.92, which is close to those for European and North American markets. 
Notably, the ApEn of the Korean foreign exchange market increased sharply from 0.55 to 1.71 after the Asian currency crisis. 
This may be attributed to the factors such as the higher volatility and the less liberalization of the Korean foreign exchange market, the stagnancy of business activities, 
and the coherent movement patterns among the companies during the Asian currency crisis. 
Our findings suggest that the ApEn can be a good measure of the market efficiency. 
\section{CONCLUSIONS}

In this paper, we have investigated the degree of randomness in the time series of various foreign exchange markets. 
We employed the Apron to quantify a market efficiency in the foreign exchange markets. 
We found that the average Apron values for European and North American foreign exchange markets are larger than those for African and Asian ones except Japan, 
indicating a higher market efficiency for European and North American foreign exchange 
markets than other foreign exchange markets. We found that the efficiency of markets with a small liquidity such as Asian foreign exchange markets 
improved significantly after the Asian currency crisis. 
Our analysis can be extended to other global financial markets.

\begin{acknowledgements}
This work was supported by a grant from the MOST/KOSHER to the National Core Research Center for Systems Bio-Dynamics (R15-2004-033), 
and by the Korea Research Foundation (KTF-2005-042-B00075), and by the Ministry of Science \& Technology through the National Research Laboratory Project, 
and by the Ministry of Education through the program KB 21, and by the Korea Research Foundation (KRF-2004-041-B00219).

\end{acknowledgements}

\end{document}